\documentclass[]{aa} 
\usepackage{graphics,txfonts,epsf,rotating,natbib,color}

\bibpunct{(}{)}{;}{a}{}{,}

\begin{document}

\title{MOST\thanks{Based on data from the {\sc Most} satellite, a Canadian Space Agency mission, jointly operated by Dynacon Inc., the University of Toronto Institute of
Aerospace Studies and the University of British Columbia with the assistance of the University of Vienna.} photometry and modeling of the rapidly oscillating \\ (roAp) star $\gamma$ Equ
}
\author{
		Michael Gruberbauer\inst{1}
		\and Hideyuki Saio\inst{2}
		\and Daniel Huber\inst{1}
		\and Thomas Kallinger\inst{1} 
		\and Werner W. Weiss\inst{1} 
		\and David B. Guenther\inst{3}
		\and Rainer Kuschnig\inst{4}
		\and Jaymie M. Matthews\inst{4}
		\and Anthony F.J. Moffat\inst{5}
		\and Slavek Rucinski\inst{6}
		\and Dimitar Sasselov\inst{7} 
		\and \\ Gordon A.H. Walker\inst{4}
		}
\titlerunning{Asteroseismology of $\gamma$ Equ}
\authorrunning{M. Gruberbauer et al.}
\offprints{M. Gruberbauer : \\ \email{gruberbauer@astro.univie.ac.at}}
\institute{
			Institute for Astronomy (IfA), University of Vienna,
	T\"urkenschanzstrasse 17, A-1180 Vienna, \\
			\email{last~name\,@\,astro.univie.ac.at}
	\and Astronomical Institute, Graduate School of Science, Tohoku University, Sendai, 980-8578, Japan, \\
			\email{saio\,@\,astr.tohoku.ac.jp}
	\and Department of Astronomy and Physics, St. Mary's University Halifax, NS B3H 3C3, Canada \\
			\email{guenther@ap.stmarys.ca}
	\and Dept. Physics and Astronomy, University of British Columbia, 6224 Agricultural Road, Vancouver, BC V6T 1Z1, Canada \\
			\email{gordonwa@uvic.ca, kuschnig@astro.phys.ubc.ca, matthews@phas.ubc.ca}
	\and D\'ept. de physique, Univ. de Montr\'eal C.P. 6128, Succ. Centre-Ville, Montr\'eal, QC H3C 3J7, Canada \\
			\email{moffat@astro.umontreal.ca}
	\and Dept. Astronomy \& Astrophysics, David Dunlop Obs., Univ. Toronto P.O. Box 360, Richmond Hill, ON L4C 4Y6, Canada \\
			\email{rucinski@astro.utoronto.ca}
	\and Harvard-Smithsonian Center for Astrophysics, 60 Garden Street, Cambridge, MA 02138, USA \\
			\email{sasselov@cfa.harvard.edu}
}	

\date{Received  / Accepted}
\abstract{}
 {Despite photometry and spectroscopy of its oscillations
obtained over the past 25 years, the pulsation frequency spectrum
of the rapidly oscillating Ap (roAp) star $\gamma$ Equ has remained poorly understood.  
Better time-series photometry, combined with recent advances to incorporate interior magnetic
field geometry into pulsational models, enable us to perform improved asteroseismology of this roAp star.}
 {We obtained 19 days of continuous high-precision photometry
of \object{$\gamma$ Equ} with the {{\sc Most}} (Microvariability \& Oscillations of
STars) satellite.  The data were reduced with two different reduction
techniques and significant frequencies were identified.  Those
frequencies were fitted by interpolating a grid of pulsation models
that include dipole magnetic fields of various polar strengths.}
 {We identify 7 frequencies in $\gamma$ Equ that we associate
with 5 high-overtone p-modes and 1st and 2nd harmonics of the dominant p-mode. 
One of the modes and both harmonics are new discoveries for this star. Our best model
solution (1.8 $M_{\sun}$, log $T_{\rm eff} \sim 3.882$; polar field
strength $\sim$ 8.1 kG) leads to unique mode identifications for
these frequencies ($\ell = $ 0, 1, 2 and 4).  This is the first
purely asteroseismic fit to a grid of magnetic models.  We measure amplitude
and phase modulation of the primary frequency due to beating with a
closely spaced frequency that had never been resolved.  This casts
doubts on theories that such modulation -- unrelated to the rotation
of the star -- is due to a stochastic excitation mechanism.}
 {}
 \keywords{stars: chemically peculiar -- stars: oscillations --  stars: individual: $\gamma$ Equ -- stars: magnetic fields -- methods: data analysis}

\maketitle

\section{Introduction - the history of $\gamma$ Equ}  
\label{intro}

{Rapidly oscillating Ap (roAp) stars form a class of variables consisting of cool magnetic Ap stars with spectral types ranging from A-F, luminosity class V, and were discovered by Kurtz (1982). Photometric and spectroscopic observations during the last 2 decades characterize roAp stars with low amplitude pulsations ($<$ 13\,mmag) and periods between 5 to 21 minutes. It is now widely accepted that the oscillations are due to high-overtone ($n >15$) non-radial p-mode pulsations. The observed frequencies can be described in a first-order approximation by the oblique pulsator model as described by \citet{kurtz82}. In this model the pulsation axis is aligned with the magnetic axis, which itself is oblique to the rotation axis of the star. It has since been refined and improved to include additional effects like the coriolis force \citep{bigotdziem02}. 

The oscillations in roAp stars are most probably excited by the $\kappa$ mechanism acting in the hydrogen ionization zone \citep{dziemgo}, while the prime candidate for the selection mechanism of the observed pulsation modes is the magnetic field \citep[e.g.][]{bigotetal00, cunhagough, balmf, bigotdziem02,  bigotdziem03, saio, cunha06}. Similar to non-oscillating Ap (noAp) stars, the group of roAp stars also shows a distinctive chemical pattern in their atmospheres, in particular an overabundance of rare earth elements, which points to stratification and vertical abundance gradients. Tackling the question of what drives the roAp pulsation, it is important to note that apart from the similar chemical composition, noAp stars show comparable rotation periods and magnetic fields but higher effective temperatures \citep{tanya04}. Thus, their hydrogen ionization zone lies further out in the stellar envelope, which prevents efficient driving of pulsation. All together, roAp stars have given new impulse to asteroseismology and 2D mapping of pulsation \citep{oleg04apj}, as well as detailed 3D mapping of (magnetic) stellar atmospheres, has become possible. 
%W+
For reviews on primarily observational aspects of roAp stars we refer to \citet{kurtz00} and \citet{kurtz04} and on theoretical aspects to \citet{shiba03} and \citet{cunha05}.

$\gamma$ Equ, (HD\,201601, HR\,8097, A9p, $m_{\rm{V}} = 4.7$) was the 6th discovery as a roAp star. 
%W-
It was long known to have a significant magnetic field \citep{babcock} which is variable with a period of $\sim$\,72 years \citep{bonsack, scholz79, leroy}.  Rotation, a solar-like magnetic cycle or precession of the star's rotation axis due to its binary companion are candidate mechanisms for this cyclic variation. The binary hypothesis was supported by \citet{scholz97}, who observed a temporary drop in radial velocity for $\gamma$ Equ during 4 consecutive nights. Their results are however contradicted by \citet{mkrtich98, mkrtich99}. Magnetic field data spanning more than 58 years indicate that the variation is most likely due to rotation and therefore $P_{\rm{mag}} = P_{\rm{rot}}$ \citep{bychkov06}. This explanation is also supported by many spectroscopic observations showing very sharp absorption lines \citep[e.g.][]{kanaan} which are typical for a slow rotator (or a star seen pole-on).

\citet{kurtz83} was the first to detect a pulsation period of 12.5\,min (1.339\,mHz) with an amplitude varying between 0.32 and 1.43\,mmag and speculated about a rotation period of 38 days, inconsistent with the magnetic field measurements mentioned above.  Aside from rotation, beating with a closely spaced frequency has been proposed as a cause, for which the present paper gives further evidence. The pulsation was confirmed shortly after Kurtz's study by \citet{weiss83}. \citet{bychkov87} reported first evidence for radial velocity (RV) variations, but it took two more years for a clear detection \citep{libbrecht}. Although unable to detect the 1.339 mHz oscillations reported by Kurtz (1983), Libbrecht discovered three frequencies at 1.365\,mHz, 1.369\,mHz and 1.427\,mHz.  He suggested that the amplitude modulation observed in the spectra of roAp stars may not be due to closely spaced frequencies, but rather caused by short mode lifetimes in the order of $\sim$ 1\,d. He concluded that both peaks therefore belong to a single p-mode oscillation. The follow-up study by \citet{weiss89} aimed at performing simultaneous spectroscopic and photometric studies but failed to confirm $\gamma$ Equ's pulsation. The latter authors applied a correlation technique using more than 100\,\AA\ wide spectral ranges in order to improve the S/N ratio relative to previous techniques, based on individual lines. This approach turned out to be inapplicable, considering the peculiar abundance structure of roAp star atmospheres which is now well known. \citet{matthews} then claimed to have detected variability with rather large radial velocity amplitudes but their results were inconsistent with previous findings. 

Detailed photometric observations were conducted by \citet{martinez} using a multi-site campaign in 1992, spanning a total of 26 nights. Their results also suggested limited life times of pulsation modes, because additional frequencies appeared in their analysis of individual nights. However, all three frequencies detected so far were confirmed, and prewhitening of 1.366\,mHz revealed a fourth eigenfrequency at 1.397\,mHz. Using these four frequencies they concluded a spacing of consecutive radial overtones of $\Delta\nu\sim 30\,\mu$Hz based on the asymptotic theory for low-degree, high-overtone p-mode pulsations \citep{tassoul}.  They did not detect any evidence for non-linear behavior of the eigenfrequencies (i.e. harmonics) which, as will be shown in this paper, is indeed present in the {\sc Most} data. So far their observing campaign was the most recent photometric investigation of $\gamma$ Equ. A period of spectroscopic studies focusing on pulsation, abundance and stratification analyses followed. 

\citet{kanaan} reported on RV amplitudes for chromium and titanium and speculated that this is due to their concentration close to the magnetic -- hence also pulsation -- poles, while other elements (e.g. iron) showing no RV variations may be concentrated at the magnetic equator. \citet{malanush} independently arrived at a similar conclusion and they are the first who identified the rare earth elements (REE) Pr{\,\sc iii} and Nd{\,\sc iii} to show the largest variations (up to 800\,$\rm ms^{-1}$). Most other atomic species had very low or non-measurable RV variations and the authors concluded that previous spectroscopic observations failed to detect significant radial velocity variations due to the chosen spectral regions and/or low spectral resolution. \citet{savanov99} further clarified the situation by drawing attention to a possible incorrect identification of spectral lines in \citet{kanaan} and concluded that Pr{\,\sc iii} and Nd{\,\sc iii} were responsible for the large RV amplitudes discussed in their analysis. 

Further investigations by \citet{oleg01} and \citet{tanya02} impressively explained amplitude modulations of different elements and even ions of the same element by lines being formed at different atmospheric depths. The authors also tried a first mode identification based on rather short data sets of line profile variations (LPVs) and they argue for $\ell = 2$ or $3$, and $m = -\ell$ or $-\ell + 1$ modes. \citet{shiba04} disagreed with this analysis and suggested a shock wave causing the observed LPVs which in turn was questioned by \citet{oleg07} who argued that the pulsational velocity should not exceed the local sound speed. The latter authors propose a modified oblique pulsator model where LPVs are caused by pulsation velocity fields superposed by sinusoidal line width changes due to convection, but which previously was thought to be suppressed. In any case, they agree with Shibahashi et al. that the identification of the primary frequency as an $\ell=1, m=0$ mode is still the most likely explanation, which is also supported by this paper. The issue is still being debated though \citep{shiba07}. 

A still unsolved issue are magnetic field variations synchronized with pulsation. \citet{leone} showed evidence for this, but subsequent investigations \citep{olegetal04, bychkov05, savanov06} could not confirm  their findings.  \citet{hubrig} corroborate this null detection for $\gamma$ Equ with ESO-FORS1 data.

Despite a history of more than 25 years of observations, the pulsation frequency spectrum of $\gamma$ Equ remains poorly understood. Several frequencies have been published, but up to now these have never been observed simultaneously. This is where {\sc Most} steps in, providing the most complete high precision photometric campaign ever conducted for $\gamma$ Equ and covering continuously a timespan of 19 days.} 

\section{Observations and data reduction}  			
\label{obsdat}

\subsection{{\sc Most} observations}						
\label{ss:most}

{{\sc Most} is a Canadian space satellite designed for the detection of stellar variability with amplitudes down to several ppm on time scales up to several days \citep{walker}. The satellite was launched in June 2003 and has since proven to deliver photometric time series of unprecedented precision. It carries a Maksutov telescope with an aperture of 15\,cm, uses a broadband filter (350--700\,nm) and operates in a low Earth orbit following the terminator in an almost polar orbit. {\sc Most} can observe stars in the  continuous viewing zone for up to 6 weeks with a pointing precision of $\sim\,1$''. An array of Fabry lenses is used to suppress effects of satellite jitter in the data of primary targets of $V<6$\,mag. $\gamma$ Equ was observed for 19 days from 07/28/2004 to 08/16/2004 with exposures (integration time $= 11$\,sec) taken continuously once every $30$\,sec. As $\gamma$ Equ was one of the first primary targets observed by {\sc Most}, stray light, produced by earthshine, and other instrumental effects were still being investigated at this time with the aim of improving the observing procedures for future runs. In fact, this data set was the basis for the development of the C reduction pipeline for {\sc Most} Fabry Imaging targets \citep{piet06}.}

\subsection{Data reduction using decorrelation}		
\label{ss:Cpipe}

{Stray light effects are corrected by computing a correlation between target and background pixels distinguished by a fixed aperture, where {\it target} denotes pixels inside the aperture and {\it background} characterizes pixels outside the aperture, which are assumed to contain no stellar signal at all. A ``neutral'' area flag, for pixels not used for either purpose, is meant to exclude pixels at the transition zone close to the edge of the aperture. 

As it turned out throughout the {\sc Most} mission, each target star's photometry suffered specific problems which needed to be addressed individually by optimizing the reduction parameters. Due to bad data quality at the beginning of the $\gamma$ Equ-run about 1060 frames had to be rejected. Additional frames needed to be eliminated due to pointing problems in the early days of {\sc Most} operations which resulted in distorted Fabry image geometries identified by comparison to a mean normalized Fabry image. 
If the pixel values in an individual (normalized) frame deviate by more than $g\,\sigma$, where $g$\, is an integer, from the pixel values of the mean normalized reference image, the frame is rejected. Since the mean reference image changes after the elimination process, this step is repeated iteratively until no more frames need to be rejected. Statistics of the reduction process are shown in Table\,\ref{tab:stat}.
%TABLE 1

\begin{table} 
\caption{Reduction statistics for $\gamma$ Equ. A total of 49213 frames were gathered. ACS: {\sc Most} Attitude Control System.}
\begin{center}
\begin{tabular}{l c c}
\hline
Reduction step & \# rejected & \% rejected \\
\hline
saturated pixels & 1 & 0.002 \\
%\hline
saturated ACS errors & 1637 & 3.33 \\
%\hline
deviating image geometry & 21679 & 44.35 \\
%\hline
excessive number of cosmics & 1002 & 3.87 \\
\hline
\end{tabular}
\label{tab:stat}
\end{center}
\end{table}

Tests showed that the number of images in which pixels are deviating by more than 5\,$\sigma$ first decreased as expected, but increased after about ten iterations (see Fig.\,\ref{fig:elim}, upper panel) and later decreased again, indicating two separate causes for deviating Fabry image geometry. Fig.\,\ref{fig:raw_phased} shows the raw light curve after correcting for cosmic rays, phased with the orbital period, clearly indicating which phases are responsible for the image geometry deviations. Not surprisingly, the majority of these rejections concern exposures taken during high stray light phases and therefore cause regular gaps in the final data set (see Fig.\,\ref{fig:lc}). To ensure that this reduced duty cycle does not impair the identification of intrinsic frequencies, the analysis was repeated with the data set obtained after the first iteration step in the image geometry evaluation, which eliminated only 807 frames.  In the lower panel of Fig.\,\ref{fig:elim} we compare the spectral windows of both reductions, showing that the rejections mainly affected orbit-induced artifacts ($f_{\rm{orbit}} = 14.19$\,d$^{-1}$).  The final analysis did not yield a different set of intrinsic, but considerably fewer instrumental frequencies.

\begin{figure}
\resizebox{\hsize}{!}{\includegraphics*{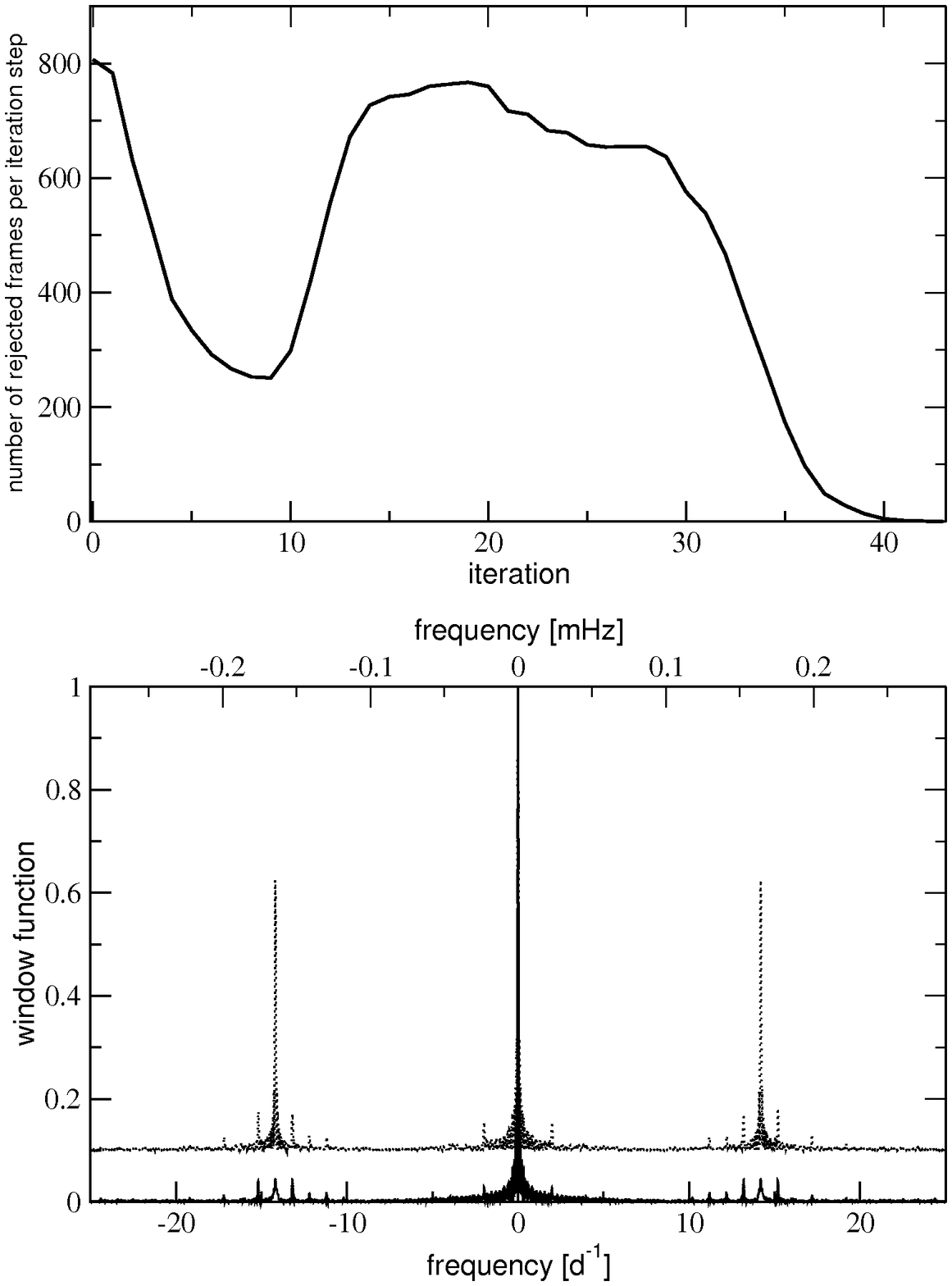}}
\caption{{\it Upper panel\,:} Iterative elimination of frames with deviating image geometry. The increasing number of rejections with higher iterations followed by a second decline indicates two separate mechanisms responsible for distorting the Fabry image geometry. \newline {\it Lower panel \,:} Window function of the data sets including (solid line) and excluding (dotted line) frames with deviating image geometry. The dotted graph has been shifted by 0.1 for better visibility.}
\label{fig:elim}
\end{figure}

\begin{figure}
\resizebox{\hsize}{!}{\includegraphics*{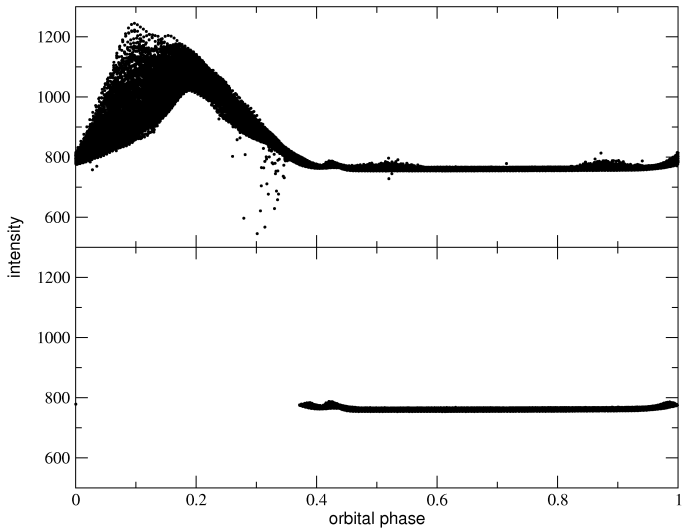}}
 \caption{{\it Upper panel\,:} The raw light curve after cosmic ray correction, phased with the {\sc Most} orbital period. The stray light contamination is mostly confined to orbital phases 0 - 0.4. \newline {\it Lower panel :} Same as in the upper panel, but after correcting for deviating image geometry. }
\label{fig:raw_phased}
\end{figure}

\begin{figure}
\resizebox{\hsize}{!}{\includegraphics*{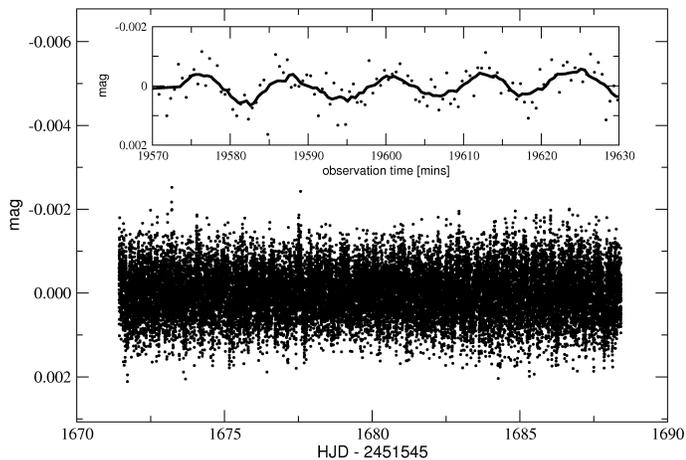}}
 \caption{Final light curve using decorrelation resulting in a duty cycle of $\sim$ 53\%. All long-term variations with periods $> 1$\,d have been prewhitened. The inset shows a subset spanning about 50\,min together with a 10 point ($\sim$\,4\,min) running average. The $\sim$\,12\,min oscillation is clearly visible.}
\label{fig:lc}
\end{figure}
}

\subsection{Data reduction using ``doughnut'' fitting}	
\label{ss:donut}
{Another (less invasive) reduction method was developed to verify the consistency of our frequency analysis. In a first step an average Fabry frame is determined from all frames obtained during orbital phases with very low stray light signal. The mean intensity, derived from pixels which are defined as background (``sky'') pixels, is subtracted from all pixels. The resulting frame is scaled to the mean intensity of the $N$ pixels with the highest intensity values, where $N$ is of the order of 100. The resulting frame serves as sort of normalized point spread function (PSF) for a Fabry image ({\sc Most}'s Fabry lenses produce images of the entrance aperture of the optics which resemble a doughnut - hence the name ``doughnut fitting''). As a next step a linear regression between pixel intensities of an image and the corresponding pixel intensities of the mean PSF frame is performed. For each of these images the slope $k$ of the linear fit gives a scaling factor, while the offset $d$ corresponds to the image mean background intensity. Each frame is replaced by the PSF frame multiplied with the corresponding scaling factor $k$ and the mean background intensity $d$ is added. The average intensities of all target and background pixels result in the target and background light curve, respectively. Finally, a linear fit between the background and target light curve is determined and subtracted from the target light curve resulting in the final light curve. In the case of $\gamma$ Equ the latter consists of 48955 datapoints. 

The advantage of this method is the insensitivity of the linear regression to extreme pixel values, for instance due to cosmic ray hits or local stray light effects. Also, contrary to the method described in section \ref{ss:Cpipe} \citep[see also][]{piet06}, the data set is not split into subsets and processed individually, which could distort the low frequency signal. The entire data set is reduced en-block, no subsets are created which qualifies this procedure as a perfect cross-check for detecting artifacts due to data reduction. The disadvantage of this method is the stray light correction which is less efficient as for the decorrelation method.} 

\section{Frequency analysis}				
\label{freqana}

{Time series resulting from both reductions outlined in \mbox{section \ref{obsdat}} were analyzed individually. {\sc SigSpec} \citep{piet07} was used to obtain frequencies by means of the spectral significance between 0 and 360\,d$^{-1}$ in a prewhitening sequence down to a significance level of 5.46, corresponding roughly to an amplitude S/N ratio of 4. Only frequencies without a counterpart in the sky background were considered, following the procedure described by \citet{pietprep}.  
{\sc Most} background light curves typically contain several dozen frequencies between 0 and 360\,d$^{-1}$, produced by stray light and instrumental effects. These are, in most cases well confined to the orbital frequency ($f_{\rm{orbit}}=14.19$\,d$^{-1}$) and harmonics of it, to 1d$^{-1}$ side lobes due to the passages of {\sc Most} above nearly the same ground pattern after 14 orbits, or are caused by temperature drifts of electronic boards. In the case of $\gamma$ Equ we used a method similar to what is described in \citet{gruberb} to identify non-intrinsic frequencies. Finally, the remaining frequencies deduced from both reduction methods were compared and, again, only matching frequencies were taken into account for our final analysis (Fig. \ref{fig:spec}). 

For the early {\sc Most} runs the low-frequency region is known to be affected by an instrumental effect producing spectral features centered on 3.16\,d$^{-1}$ and multiples of it \citep{piet06}. This instrumental effect is much weaker for the background  pixels which explains why the corresponding frequencies remain even after comparisons with the background frequency spectrum and what justifies the rejection of all frequencies below 10\,d$^{-1}$. Some of the power excess in the low frequency region might be still of stellar origin, but the current data is inconclusive. We also found the frequencies between 10 and 35\,d$^{-1}$ to be multiples of 3.16\,d$^{-1}$, but no other {\sc Most} data set shows these instrumental effects at such high frequencies. Because we prefer a critical approach, in the end, only the six higher frequencies with $f >100$\ d$^{-1}$ were considered to be intrinsic to $\gamma$ Equ beyond doubt. 

Table\,\ref{tab:freqlist} lists the result of our conservative analysis. Two additional significant frequencies, not included in the table, can be found very close to $f_{\rm{1}}$ with amplitudes of $\sim 40$\,ppm, but including them in a Period04 multi-sine fit \citep{lenz} fails, because the solution does not converge. Their nature is discussed in section \ref{ss:ampmod}. 

\begin{table}
\caption{The list of frequencies considered to be intrinsic to $\gamma$ Equ. {$\sigma_f$:} 1\,$\sigma$--uncertainty in frequency, derived according to \citet{kallinger}; {$\sigma_a$:} uncertainty in amplitude; {\sc sig}: spectral significance according to \citet{piet07}; {$\theta$:} phase in radians corresponding to $cos(2\pi f_i t - \theta )$.}
\begin{center}
\begin{tabular}{c c c c c c c}
\hline 
\hline
      & {$f$}&{$\sigma_f \cdot 10^6$} &{$a$} &{$\sigma_a$} &{$\sc sig$}& {$\theta$} \\
      & mHz &  mHz & ppm & ppm & & \\
\hline 
$f_{\rm{1}}$ & 1.364594 &  6  & 362 & 2.9 & 963 & 2.7704 \\
$f_{\rm{2}}$ & 1.365411 & 13  & 124 & 2.4 & 175 & 2.6594 \\
$f_{\rm{3}}$ & 1.427102 & 34  &  43 & 2.1 &  25 & 1.8325 \\
$f_{\rm{4}}$ & 1.388872 & 44  &  32 & 2.1 &  15 & 1.3636 \\
$f_{\rm{5}}$ & 1.310914 & 51  &  27 & 2.1 &  11 & 2.7173 \\
$f_{\rm{6}}$ = 2$f_{\rm{1}}$ & 2.729148 & 32  &  44 & 2.1 &  27 & 2.0870 \\
$f_{\rm{7}}$ = 3$f_{\rm{1}}$ & 4.094687 & 70 &  18 & 3.7 &    6 &  3.1012 \\
\hline
\end{tabular}
\label{tab:freqlist}
\end{center}
\end{table}

\begin{figure} 
\begin{center}
\resizebox{\hsize}{!}{\includegraphics*{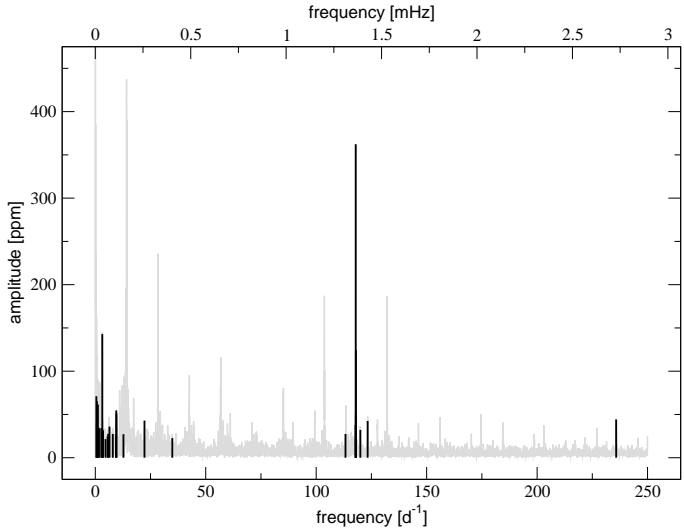}}
\caption{The amplitude spectrum of the final light curve ({\it gray}) presented in Fig.\,\ref{fig:lc} and all frequencies found by {\sc SigSpec} in both reduction methods described in this paper ({\it black}). We consider only frequencies with $f > 50$\,d$^{-1}$ to be unaffected by the instrument. $\gamma$ Equ's pulsation is obvious in the range of 100 to 125\,d$^{-1}$ and the first of two harmonics of $f_{\rm{1}}$ is evident at $\sim 235$\,d$^{-1}$. }
\label{fig:spec}
\end{center}
\end{figure}

No obvious spacing, as predicted by the asymptotic theory of non-radial pulsation, can be found in our final set of frequencies. This suggests that the regular spacing of modes with the same degree $\ell$ is disturbed, e.g. by the magnetic field, or that modes of different degree are excited. We therefore turned to fit our observed frequencies to the latest generation of roAp-pulsation models, but using only $f_{\rm{1}}$ to $f_{\rm{5}}$ for the fitting process, as $f_{\rm{6}}$ and $f_{\rm{7}}$ correspond to the harmonics of $f_{\rm{1}}$.}

\section{Pulsation models}
\label{pulsmod}

%%%%%%%%%%%%%%%%%%%%%%%%%%%%%%%%%%%%%%%%%%%%%%%%%%%%%%%%%%%%%%%%%%%%%
{To compare theoretical frequencies with observed ones of \mbox{$\gamma$ Equ}, we have computed main-sequence evolutionary models for a mass range of $1.75\,M_{\rm{\odot}} - 1.85\,M_{\rm{\odot}}$ with heavy element abundance of $0.015 \le Z \le 0.025$ and OPAL opacities \citep{OPAL}. A list of the individual models, and how we organised them into grids, is presented in Table\,\ref{tab:grids}. For standard models, envelope convection is suppressed ($\alpha = 0.0$), and to mimic depletion of helium abundance in the outermost layers of a star, the helium abundance $Y$ is given as $Y = 0.01 + 0.27(x_2+x_3)$, where $x_2$ and $x_3$ are fractions of singly and doubly ionized helium, respectively \citep[cf][]{balmf}. Nonetheless, the effects of convection were also tested by computing all grids except {\it Grid 4} with $\alpha = 1.5$ and homogeneous He abundance in the outer layers, but otherwise the same parameters. {\it Grid 4} included a homogeneous distribution (mixing) of He in the envelope but no convection.

Nonadiabatic frequencies of axisymmetric ($m=0$) high order p-modes under the presence of a dipole magnetic field were calculated using the method described in \citet{saio}. Outer boundary conditions are imposed at an optical depth of $10^{-3}$. A reflective mechanical condition ($\delta p/p \rightarrow $ constant) is adopted. All frequencies presented in this paper are less than the acoustic critical frequencies of corresponding models. 

Since the latitudinal dependence of amplitude under the presence of a magnetic field cannot be expressed by a single Legendre function, we expand it into a truncated series of components proportional to Legendre functions $P_{l_j}$ with $l_j = 2j - 1$ for odd modes and $l_j = 2j$ for even modes, where we have included twelve terms; i.e., $j = 1, 2, \ldots, 12$. The latitudinal degree $\ell$ is not a definite quantity for a pulsation mode any more, because pulsation energy is distributed among twelve components associated with $P_{l_j}$. For convenience we still use $\ell$ representing the $l_j$ value of the component associated with the largest kinetic energy. Sometimes the distribution of the kinetic energy among the components is broad so that the identification of $\ell$ is ambiguous, and the value of $\ell$ may change as the strength of the magnetic field changes. Also, for a mode associated with $\ell$, the latitudinal dependence of amplitude on the stellar surface may be considerably different from that of $P_\ell(\cos\theta)$ as shown in \citet{saiogau, saio} (see also below).}
%%%%%%%%%%%%%%%%%%%%%%%%%%%%%%%%%%%%%%%%%%%%%%%%%%%%%%%%%%%%%%%%%%%%%%

\section{Model fitting}
\label{pulsfit}

{Pulsation models were calculated taking into account a magnetic dipole with polar field strength, $B_{\rm{P}}$, up to 12\,kG. These models were tried to match the currently estimated parameter space of $\gamma$ Equ, which is $\log T_{\rm{eff}} = 3.882 \pm 0.011$\,K, $\log L/L_{\rm{\odot}} = 1.10 \pm 0.03$, and $M \sim 1.74 \pm 0.03$\,M$_{\rm{\odot}}$ \citep{olegbagnulo}. Fig.\,\ref{fig:HRDpos} shows the position and parameters of all models involved. For the corresponding values and further details, we again refer to Table\,\ref{tab:grids}.

To increase the resolution of our model grid, the mode frequencies were interpolated linearly in $B_{\rm{P}}$ and ($\log T_{\rm{eff}}$, $\log L/L_{\rm{\odot}}$)  for a fixed mass and chemical composition. The step width for the interpolation was chosen to be 0.02\,kG in $B_{\rm{P}}$ and 0.00002 in $\log T_{\rm{eff}}$. Since not all modes with the same degree $\ell$ and radial order $n$ converge in the model calculations, the sequence of frequencies for a specific mode may not cover the entire chosen stellar fundamental parameter space ($\log T_{\rm{eff}}, \log L/L_{\rm{\odot}}$, and $M$). Finally we used a single coordinate, $\Delta \nu$, according to 

\begin{equation}
\label{equ:deltaNu}
\Delta\nu = 0.1349 \cdot\left(\frac{\rho}{\rho_\odot}\right)^{0.5},
\end{equation}

\noindent where $\rho$ is the mean density of the stellar model. All model fitting was carried out for a given mass in the ($B_{\rm{P}}$,$\Delta\nu$)--space with a $\chi^2$--test similar to \citet{guenther}. Since an {\it a priori} mode identification was not a reasonable option, each model's eigenfrequencies of spherical degrees $\ell = 0$ to $\ell = 4$ were compared to the five observed frequencies. The 1\,$\sigma$--uncertainties of the observed frequencies were estimated to be roughly 0.25 of the upper frequency error derived according to \citet{kallinger}. The corresponding 1\,$\sigma$--errors assigned to the individual model frequencies were assumed to be 0.2\,$\mu$Hz.

The best fitted models, represented by a minimum $\chi^2$, are listed in Table\,\ref{tab:fitres}. {\it Grid 3}, as defined in Table\,\ref{tab:grids}, outperforms the other models by far. It is the only model grid that manages to produce frequencies that fit all five observed frequencies on average to within the uncertainties ($\chi^2 \le 1$), as it is illustrated by the upper panel of Fig.\,\ref{fig:chisquare}. As expected, including convection for {\it Grid 3} destroys the fit, while for grids with large $\chi^2$ it has the opposite effect. The best fit is located at the center of an extended patch. Its proximity to a genuine calculated model ensures that this good fit is not an artifact produced by the interpolation (also see Section\,\ref{discuss}). The mean parameter values for the $\chi^2 \le 1$ region with the mentioned interpolation step size are $\log T_{\rm{eff}}=3.8818$, $\log L/L_{\rm{\odot}} = 1.0871$ and $B_{\rm{P}} = 8.1\,\rm{kG}$.
 
Fig.\,\ref{fig:bestfit} shows an \'echelle diagram of the best fitted model together with the observations. $f_{\rm{1}}$, the primary frequency, is identified as an $\ell=1$ mode. $f_{\rm{2}}$ and $f_{\rm{3}}$ are matched by consecutive $\ell=4$ modes. The remaining two frequencies, $f_{\rm{4}}$ and $f_{\rm{5}}$, are fitted by modes of degree $\ell=2$ and $\ell=0$. As a comparison, the lower panel of Fig.\,\ref{fig:chisquare} shows the fitting results to {\it Grid 6}, a grid with considerably larger $\chi^2$ values. Interestingly, while all grids except {\it Grid 3} fail at delivering $\chi^2 < 1$--results, the mode identification remains stable for grids with $\chi^2 < 4$. In particular, the two closely spaced model frequencies in the vicinity of $f_{\rm{1}}$ and $f_{\rm{2}}$ are found to fit the same degrees $\ell$\,=\,1 and 4 for {\it Grids 3}, {\it 7$\alpha$}, {5$\alpha$}, and {\it 6}. This is not the case for {\it 6$\alpha$} and {\it 1$\alpha$}. Thus, in our grids there seems to be a tendency towards the mode identification of {\it Grid 3} for better fitted models. Also, convection is shown to radically influence the model frequencies. 

In Fig.\,\ref{fig:chisquare} obvious discontinuities can be seen in the fitting results. They are produced when the interpolation routine cannot find two modes of equal spherical degree $\ell$ and radial order $n$
in both models that are acting as sampling points. This can happen, as already mentioned in Section\,\ref{pulsmod}, when the $\ell$-value of a mode formally changes, because the former associated spherical harmonic does no longer supply most of the kinetic energy. Also, the cyclic variation of the damping rate as a function of $B_{\rm{P}}$, as shown in \citet{saiogau}, has an effect on the ``availability'' of certain modes. Around the maximum of the damping rate, the kinetic energy is so broadly distributed among the different $\ell$-components that convergence of the model calculations starts to fail. This effect is clearly visible as the two large discontinuities in each panel of Fig.\,\ref{fig:chisquare}, which bears resemblance to what is presented in \citet{saiogau}.   

\begin{table}
\caption{List of all stellar evolution models computed for $\gamma$ Equ with polar magnetic field strengths ranging from 0 to 10\,kG in steps of 0.2\,kG, except of {\it Grid 7} which was calculated up to a magnetic field strength of 12\,kG.}
\begin{center}
\begin{tabular}{c c c c c}
\hline
\hline
{Grid ID} & {$M/M_{\rm{\odot}}$} & {Z} &  {$\log L/L_{\rm{\odot}}$} & {$\log T_{\rm{eff}}$}  \\
\hline 
 1a & 1.75  &  0.02  & 1.0252 & 3.8810 \\
 1b  & 1.75  &  0.02  & 1.0229 & 3.8825 \\
 \hline 
 2a   & 1.80  &  0.015 &  1.1574 & 3.8990  \\
 2b & 1.80  &  0.015 &  1.1558 & 3.9004  \\
 2c   & 1.80  &  0.015 &  1.1542 & 3.9017  \\
 \hline
 3a   & 1.80  &  0.02 &  1.0880 & 3.8809  \\
 3b  & 1.80  &  0.02 &  1.0871 & 3.8818  \\
 3c   & 1.80  &  0.02 &  1.0861 & 3.8826  \\
 \hline 
 4a  &  1.80 & 0.02* & 1.0899 & 3.8794 \\
 4b  &  1.80 & 0.02* & 1.0880 & 3.8810 \\ 
 4c  &  1.80 & 0.02* & 1.0871 & 3.8819 \\
 4d  &  1.80 & 0.02* & 1.0861 & 3.8827 \\
 \hline
 5a   & 1.80  &  0.025 &  1.0284 & 3.8701  \\
 5b  & 1.80  &  0.025 &  1.0267 & 3.8714  \\
 5c   & 1.80  &  0.025 &  1.0243 & 3.8726  \\
\hline 
 6a  & 1.85  &  0.02 &  1.1388 & 3.8885 \\
 6b  & 1.85  &  0.02 &  1.1368 & 3.8903 \\
 6c  & 1.85  &  0.02 &  1.1345 & 3.8917  \\
 6d  & 1.85  &  0.02 &  1.1325  & 3.8934  \\
\hline 
7a & 1.85  &  0.025 &  1.0753 & 3.8806 \\
7b & 1.85  &  0.025 &  1.0734 & 3.8819 \\
7c & 1.85  &  0.025 &  1.0715 & 3.8832  \\
\hline
\end{tabular}
\label{tab:grids}
\end{center}
\end{table}

\begin{figure} 
\begin{center}
\resizebox{\hsize}{!}{\includegraphics*{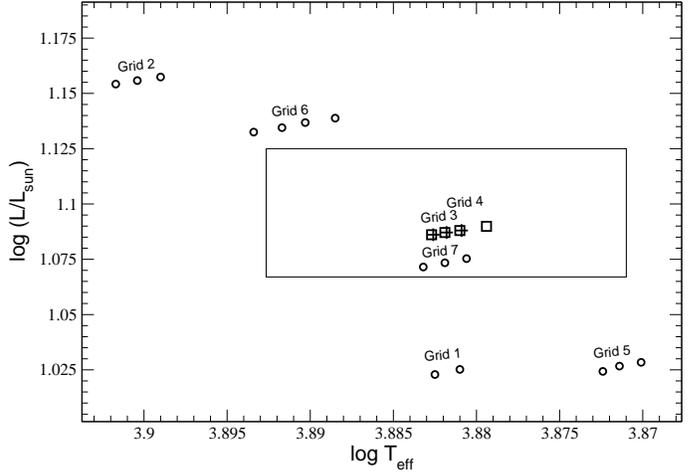}}
\caption{A schematic diagram of the calculated grids and their model parameters following the notation of Table\,\ref{tab:grids}. Since {\it Grid 3} and {\it Grid 4} partially overlap, all grid points for {\it Grid 3} are shown as + symbols, while {\it Grid 4} is indicated by rectangles. Observational uncertainties for $\gamma$ Equ's estimated position in the HR diagram \citep{olegbagnulo} are represented by the large rectangle.}
\label{fig:HRDpos}
\end{center}
\end{figure}

\begin{figure} 
\resizebox{\hsize}{!}{\includegraphics*{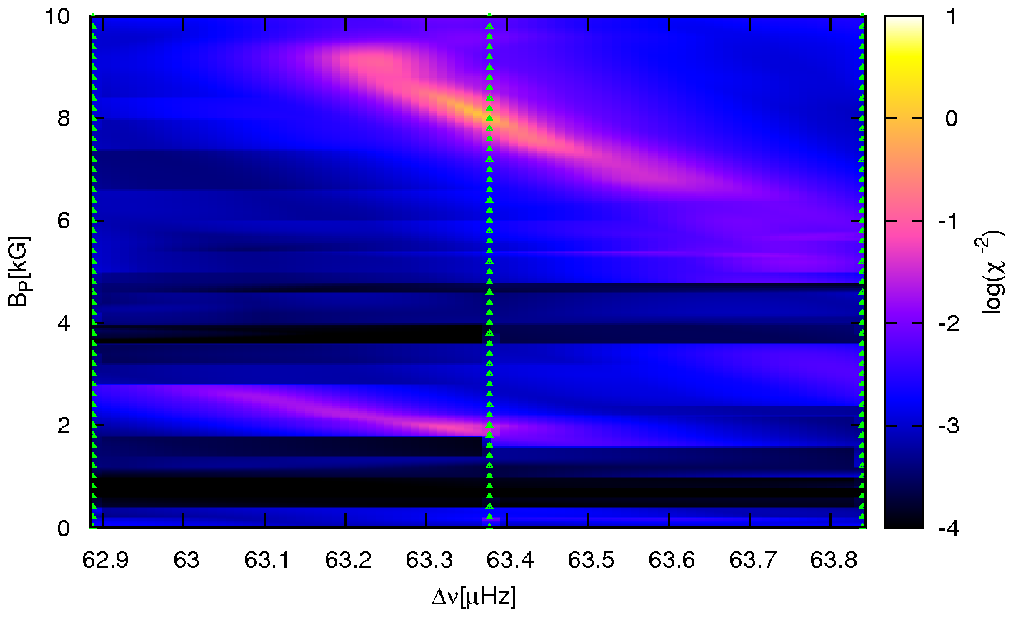}}
\resizebox{\hsize}{!}{\includegraphics*{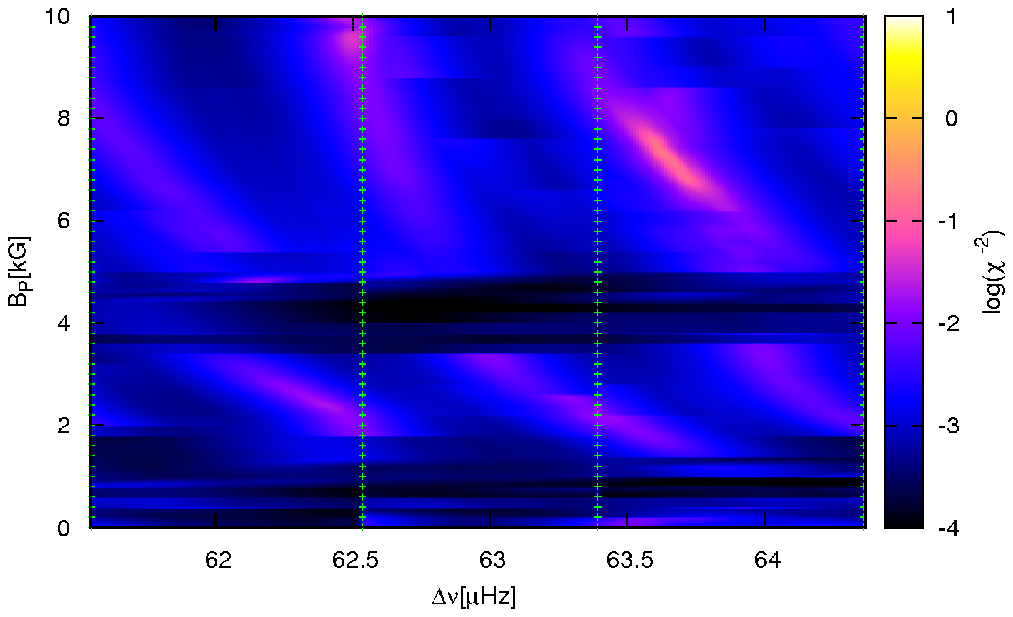}}
\caption{{\it Upper panel:} Color coded inverse $\chi^2$ values obtained when fitting the 5 intrinsic frequencies of $\gamma$ Equ to model frequencies of {\it Grid 3}. The vertically aligned symbols represent the calculated model positions -- all other model frequencies are derived via linear interpolation. 
Discontinuities are due to a lack of converging model frequencies. Some modes of a certain degree and radial order are missing in the individual genuine models. Also, sudden changes of the frequency spacing due to the magnetic field, can change the quality of the fit quite rapidly. \newline {\it Lower panel: Grid 6} shows a pattern similar to {\it Grid 3} but delivers larger $\chi^2$ values. It gives, however, the same mode identification. Note the different scale of the abscissa!}
\label{fig:chisquare}
\end{figure}

\begin{table}
\caption{ $\chi^2$ values of the best fits in the corresponding model grids. Models taking convection into account are denoted by an index $\alpha$.}   
\begin{center}
\begin{tabular}{c c c}
\hline
\hline
{Grid ID} &  {min($\chi^2$)} & $B_{\rm{P}}\rm{[kG]}$  \\
\hline
1  &  41.0 &  6.0 \\
1$\alpha$  & 4.4 & 8.8 \\
 \hline
 2  &   6.0  & 6.5 \\
 2$\alpha$ & 21.7 & 6.5 \\
 \hline
 3  &   0.7  & 8.1 \\
 3$\alpha$  &  7.7 & 2.8 \\   
 \hline
 4  &  8.0  &  0.0 \\
 \hline
 5  &   22.6  & 0.4 \\ 
 5$\alpha$ & 2.2 &  4.6 \\
 \hline
 6  &   4.6  & 7.3 \\
 6$\alpha$ & 4.6 & 6.0 \\
 \hline
 7  &   17.8 & 0.0 \\
 7$\alpha$ & 2.0 & 3.7 \\
\hline
\end{tabular}
\label{tab:fitres}
\end{center}
\end{table}

\begin{figure}
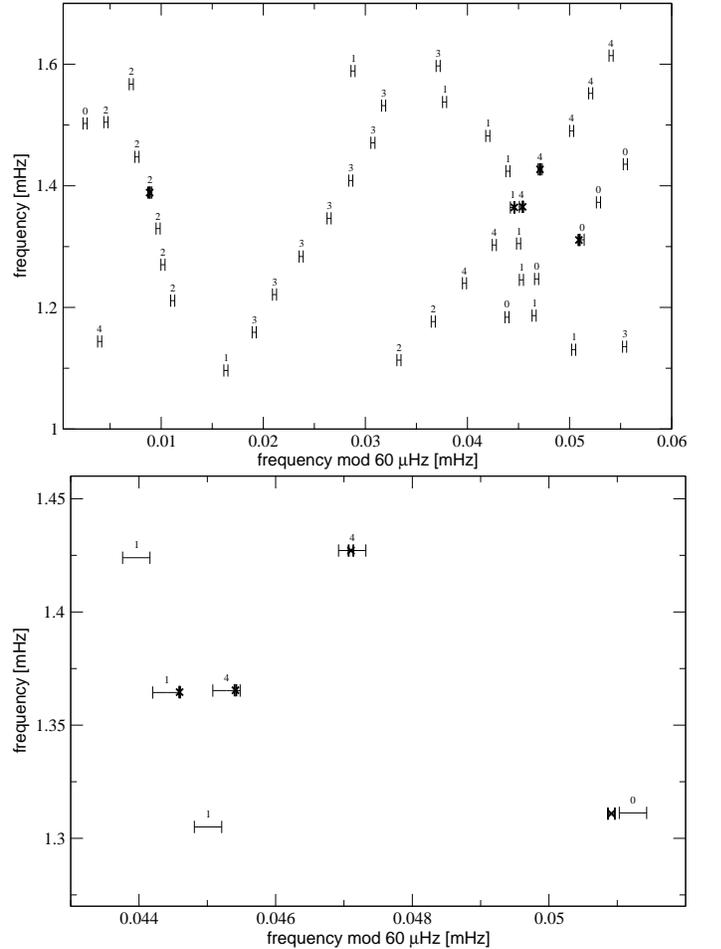
 
\begin{center}
\resizebox{\hsize}{!}{\includegraphics*{8830fg7a.eps}}
\resizebox{\hsize}{!}{\includegraphics*{8830fg7b.eps}}
\caption{{\it Upper panel: } \'Echelle diagram of the best fitted stellar model ($\chi^2 = 0.744$, {\it Grid 3}). The corresponding parameters are $\log T_{\rm{eff}}=3.8818$, $\log L/L_{\rm{\odot}}=1.0871$, and $B_{\rm{P}} = 8.06$\,kG. The model frequencies are represented by their estimated error bars, the numbers above denote the mode degree $\ell$. Black cross symbols indicate the position of the observed frequencies with error bars. \newline {\it Lower panel: } A close-up of the region in the \'echelle diagram including $f_{\rm{1}}$, $f_{\rm{2}}$, $f_{\rm{3}}$, and $f_{\rm{5}}$ illustrating the quality of the fit more clearly. $f_{\rm{5}}$ is not matched to within the 1\,$\sigma$--error, but $f_{\rm{1}}$ to $f_{\rm{3}}$ dominate the $\chi^2$ statistics because of their high significance.}
\label{fig:bestfit}
\end{center}
\end{figure}
}

\section{Discussion}             
\label{discuss}

\subsection{The frequencies of $\gamma$ Equ}
{As mentioned in the introduction, roAp stars are characterized by an interaction of a strong global magnetic field with velocity fields due to pulsation of still poorly known origin, all leading to a complex eigenfrequency spectrum. The latter provide the only directly accessible information concerning the internal structure of these stars.

Data obtained by {\sc Most} have allowed us to unambiguously identify 7 frequencies (Table\,\ref{tab:freqlist}) above 1.3\,mHz, including the first two harmonics of the primary frequency, which significantly exceed the mean noise level of $\sim 10$\,ppm between 1.15 and 1.75\,mHz. The comparison with Martinez et al. yields a match for $f_{\rm{1}}$, $f_{\rm{3}}$, and $f_{\rm{4}}$. Our $f_{\rm{2}}$ has never been detected before, probably because $f_{\rm{1}}$ and $f_{\rm{2}}$ could not be resolved as individual frequencies. This is also important for the discussion on amplitude modulation of the primary frequency, for which we refer to the next section. Our value for $f_{\rm{5}}$ is comparable to their $\nu_1=1.321$\,mHz (taken from Weiss 1983), but differs by $\sim 0.01$\,mHz.  We assume that they have misidentified a 1-day alias of the real frequency but this is difficult to asses, since no frequency uncertainties are known for this data set. Given that the data set from \citet{weiss83}  spans only three nights, the poor frequency resolution $1/T_{\rm{obs}} \simeq 0.33\,\rm{d}^{-1} = 0.004$\,mHz supports our assumption. Thus, we can for the first time confirm and expand the previous set of frequencies unambiguously.

When discussing published amplitudes (including those in the present paper) one has to keep in mind that two different properties prevent a direct comparison. First, amplitude modulations are present which lead to different amplitudes for observations obtained at different beating phases. Second, it has been well known since the eighties that amplitudes \citep{weiss84} and  phases \citep{weiss86} depend on wavelength and the former are rapidly decreasing towards the red. The passband of {\sc Most} is broad compared to the Str\"omgren system, frequently used for roAp star photometry, hence the observed {\sc Most} amplitudes are intrinsically smaller.}

\subsection{Amplitude modulation}
\label{ss:ampmod}
{Amplitude modulation of $\gamma$ Equ's primary frequency has often been mentioned in the literature and rotation or closely spaced frequencies have been proposed as possible explanations. Because of the relatively small effect and the limited accuracy of the data a specific modulation period could never be accurately determined. This situation, however, changed with {\sc Most}. To distinguish between amplitude modulation of a single frequency and beating of two closely spaced frequencies one needs to discuss simultaneous phase $and$ amplitude changes \citep{breger06}. Figure\,\ref{fig:ampmod} illustrates the correlation of amplitude and phase variations for $f_{\rm{1}}$ with a relative phase shift of $\frac{\pi}{2}$, which is indicative for beating of a close pair of frequencies spaced by $f_{\rm{beat}} \sim 0.07$\,d$^{-1}$~~($\sim 0.0008$\,mHz). 

We find $f_{\rm{1}} + f_{\rm{beat}} \simeq f_{\rm{2}}$. The best fitting pulsation model predicts both $f_{\rm{1}}$ and $f_{\rm{2}}$ to be eigenfrequencies of the star, hence we conclude that the ``closely spaced frequencies''-hypothesis is correct. It comes as no surprise that previous observations, especially time-resolved spectroscopy, could not explain the modulation effect, since it is impossible to resolve $f_{\rm{2}}$ with short time bases. The possibility of unresolved modes in such data should be taken into account when studying pulsation via residual spectra, produced by subtraction of a mean spectrum phased by only a single pulsation frequency. For ground-based photometry, the aliasing problem and the higher noise level might also have contributed to the difficulty of detecting $f_{\rm{2}}$. 

We have found no evidence for modulation of the other frequencies, but we can comment on two other significant frequencies with amplitudes $\sim 40$\,ppm close to  $f_{\rm{1}}$ which are not instrumental. These frequencies appear to be Fourier artifacts produced by irregularities in the beating of $f_{\rm{1}}$ and $f_{\rm{2}}$. As mentioned in Section\,\ref{freqana}, a multi-sine fit fails to converge if they are included in the solution. We have to mention here again that the broad filter band pass used by {\sc Most} may be disadvantageous for such investigations.

\begin{figure} 
\begin{center}
\resizebox{\hsize}{!}{\includegraphics*{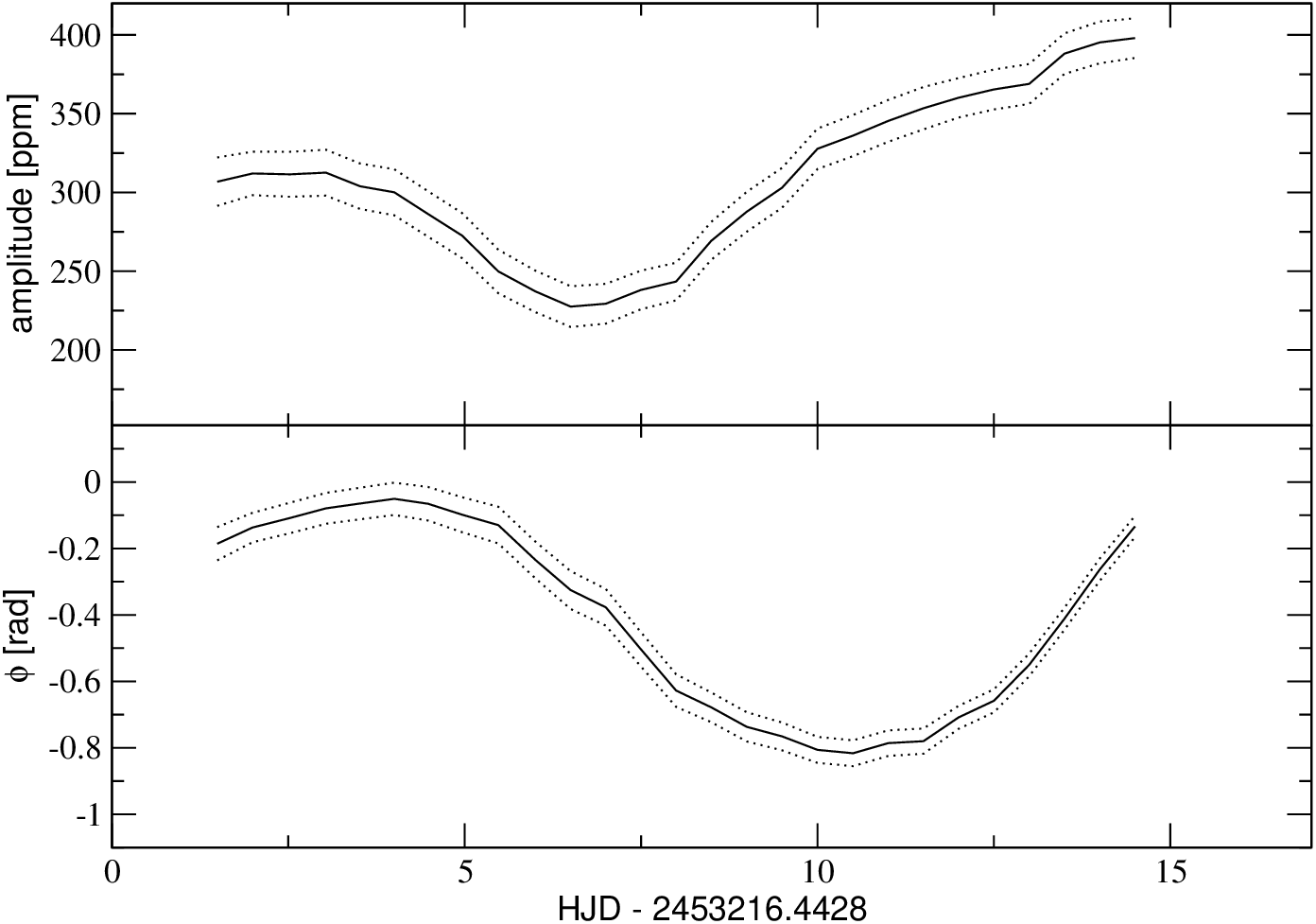}}
\caption{{\it Upper panel: }  Amplitude modulation of the primary frequency, $f_{\rm{1}}$, as found by a least-squares fit with a fixed frequency to subsets of the light curve. \newline {\it Lower panel: } Phase changes of $f_{\rm{1}}$ determined in the same way as the amplitude modulation. The dotted lines represent the 1\,$\sigma$--confidence limits. The similarity of amplitude modulation and phase shift, as well as the minimum amplitude coinciding with the time of the phase shift turning point, suggest an amplitude modulation due to beating of two closely spaced frequencies with a beat frequency of $\sim 0.07$\,d$^{-1}$ ($\sim 0.0008$\,mHz). This corresponds to the beat frequency of $f_{\rm{1}}$ and $f_{\rm{2}}$.}
\label{fig:ampmod}
\end{center}
\end{figure}
}

\subsection{On the validity of interpolated frequencies}
\label{ss:modeldisc}

{To increase the resolution of our grid, we used linear interpolation in temperature, luminosity, and polar magnetic field strength, respectively. It may be questioned if such an interpolation is a sensible approach. For a test we compared interpolated values with genuine models. For all genuine models $m_{\rm{g}}$ enclosed by 2 additional genuine models $m_{\rm{g,1}}$ and $m_{\rm{g,2}}$ , we produced frequencies at the position of $m_{\rm{g}}$ by interpolation between $m_{\rm{g,1}}$ and $m_{\rm{g,2}}$. The deviation of the interpolated frequencies from their genuine values, normalized to the gap in $\log T_{\rm{eff}}$ between the enclosing and the enclosed models, was calculated and over 16000 frequencies were compared. Figure \,\ref{fig:iphist} shows a histogram of the results. It illustrates that our interpolation delivers a reasonable approximation to genuine models, as long as the gaps in between the reference models remain reasonably small. 

In the case of {\it Grid 3} (see Table\,\ref{tab:grids}) we also repeated the frequency fitting procedure.  
We interpolated between {\it 3a} and {\it 3c} and omitted the {\it 3b}--models of {\it Grid 3} . Figure\,\ref{fig:iptest} shows the $\chi^2$--statistics for our observed frequencies based on an interpolation in this much coarser grid (best fit with $\chi^2 = 2.642$). A comparison of Figure\,\ref{fig:iptest} with the upper panel of Fig.\,\ref{fig:chisquare} again shows that interpolation is applicable for small gaps in between genuine models, which is also reflected in the results of the fitting procedure. Since not all modes can be approximated sufficiently through linear interpolation, as indicated by the outliers in Figure \,\ref{fig:iphist}, it is necessary to check whether the modeled frequencies that fit the observations satisfy the linear assumption. Figure\,\ref{fig:bmtest} presents the absolute (rather than the normalized) deviation of the interpolated frequencies from the genuine frequencies for the {\it 3b}--models. It shows that the interpolated values of all 5 modes that fit the observations are hardly deviating from their genuine values - the deviations lie well within the model uncertainties. 
 
It is important to note that all of the modes which fit the observed frequencies are actually not expected to be excited according to current model physics. This hints at remaining problems concerning the stability analysis or the implemented driving mechanism. 

\begin{figure}[h!]
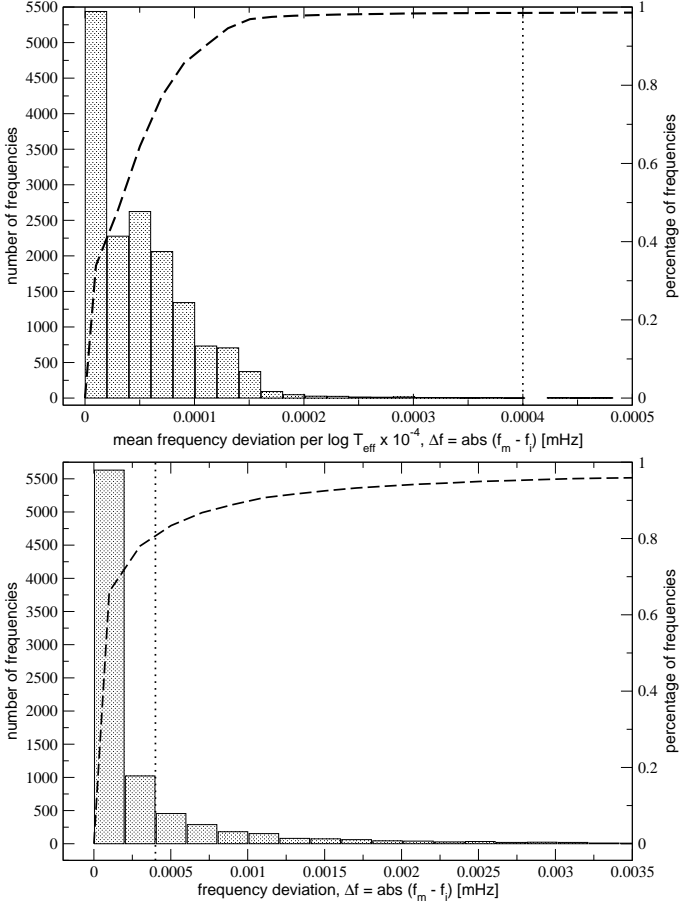
 
\begin{center}
\resizebox{\hsize}{!}{\includegraphics*{8830fg9a.eps}}
\resizebox{\hsize}{!}{\includegraphics*{8830fg9b.eps}}
\caption{{\it Upper panel: } A histogram of the normalized deviation of 16050 frequencies, interpolated as a function of $\log T_{\rm{eff}}$, from their values in genuine models. The bars show the number of frequencies deviating within a certain range. The dotted line gives the estimated upper limit of the model uncertainties. The dashed line shows the total percentage of frequencies, as indicated by the right-hand side ordinate, as a function of frequency deviation. \newline {\it Lower panel: } A histogram of the absolute frequency deviation of 8523 frequencies, interpolated as a function of $B_{\rm{P}}$, from their genuine counterparts. No normalization was necessary, since the interpolation was performed with a step size of 0.2\,kG for each frequency. The dashed and dotted lines have the same meaning as in the upper panel.}
\label{fig:iphist}
\end{center}
\end{figure}

\begin{figure}[h!] 
\begin{center}
\resizebox{\hsize}{!}{\includegraphics*{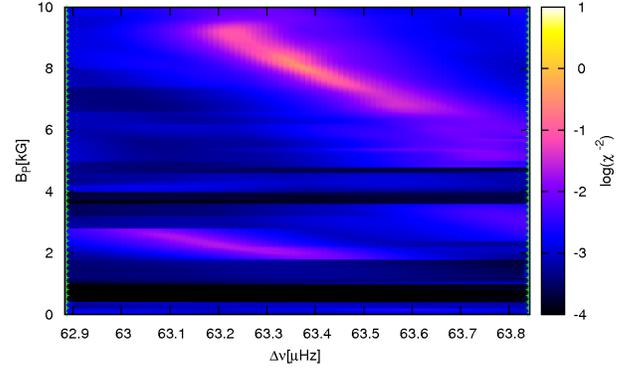}}
\caption{The result of fitting our observed frequencies to a model grid calculated by interpolation between {\it 3a} and {\it 3c}. Although the {\it 3b}--models
are omitted and no fit with $\chi^2 \le 1$ is found, the figure compares well to the upper panel of Fig.\,\ref{fig:chisquare}. Linear interpolation in $B_{\rm{P}}$ is therefore
a reasonable approximation to genuine models for our fitting procedure.}
\label{fig:iptest}
\end{center}
\end{figure}

\begin{figure}[h!] 
\begin{center}
\resizebox{\hsize}{!}{\includegraphics*{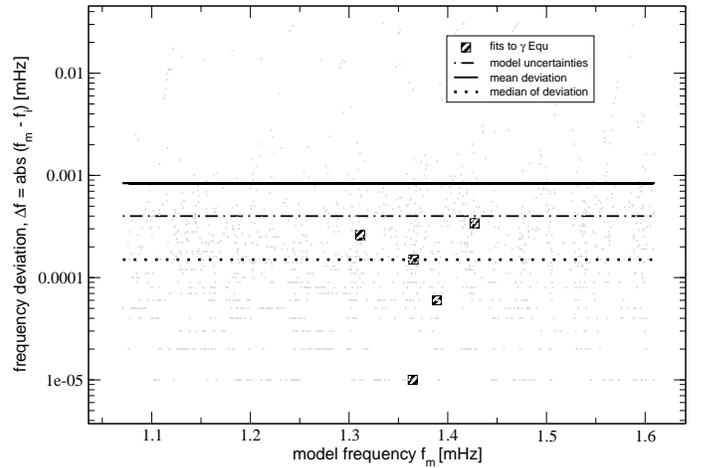}}
\caption{The absolute deviations of all interpolated frequencies at the position of {3b}, using Grid models {3a} and {3c}, from the genuine frequency values. All 5 modes, which match the observed frequencies of $\gamma$ Equ best and are indicated by boxes, are deviating less than the model uncertainties. }
\label{fig:bmtest}
\end{center}
\end{figure}
}
%%%%%%%%%%%%%%%%%%%%%%%%%%%%%%%%%%%%%%%%%%%%%%%%%%%%%%%%%%%%%%%%%%%%%%%%%%%
\subsection{Latitudinal amplitude dependence}
{As discussed in \S\ref{pulsfit} the observed frequencies of $\gamma$ Equ are identified as modes of $\ell =$ 0, 1, 2, and 4. Those modes, however, have amplitude distribution on the stellar surface considerably deviating from that of a single Legendre function $P_\ell(\cos\theta)$ due to the strong magnetic effect (Fig.\ref{fig:latamp}). The amplitude for $f_{\rm{1}}$ ($\ell=1$) and $f_{\rm{4}}$ ($\ell=2$) is more concentrated toward the polar regions than $P_1(\cos\theta)$ and $P_2(\cos\theta)$. It is interesting that the amplitude distribution of $f_{\rm{4}}$ ($\ell=2$) on a hemisphere is not very different from that of $f_{\rm{1}}$ ($\ell=1$) having very small amplitude near the equator, although $f_{\rm{1}}$ is antisymmetric and $f_{\rm{4}}$ symmetric to the equator.  

For $f_{\rm{2}}$, $f_{\rm{3}}$ ($\ell=4$) and $f_{\rm{5}}$ ($\ell=0$) the amplitude distributions are strongly concentrated around $\cos\theta \approx 0.65$, drastically different from those for non-magnetic stars. Although the light variation from an $\ell=4$ mode of a non-magnetic star is expected to suffer from a strong cancellation on the stellar disk, Fig. \ref{fig:latamp} indicates that our cases $f_{\rm{2}}$ and $f_{\rm{3}}$ seem hardly affected by this effect.  

\begin{figure}
\begin{center}
\resizebox{\hsize}{!}{\includegraphics*{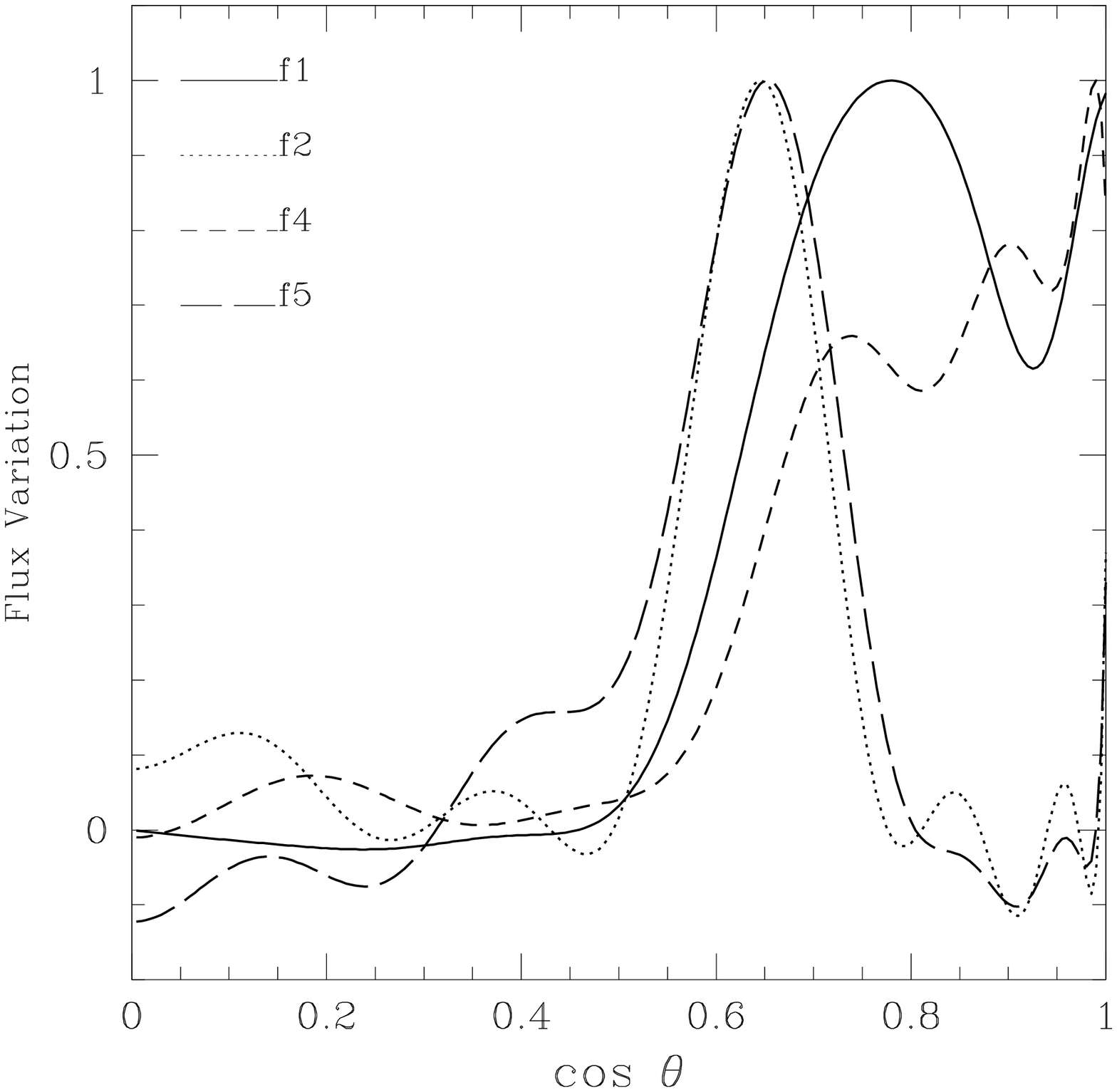}}
\caption{Latitudinal variation of the perturbation of radiative flux on the surface is shown for each mode matched to $\gamma$ Equ. (The case of $f_{\rm{3}}$ is not shown because it is very close to that of $f_{\rm{2}}$.) The ordinate shows $\cos\theta$ with $\theta$ being co-latitude ($\cos\theta = 1$ at poles). The plots are the real parts of the eigenfunctions for the flux perturbation at $B_{\rm{P}}=8$\,kG for model {\it 3b} (Table \ref{tab:grids}). Each curve is normalized as that the maximum is unity.}
\label{fig:latamp}
\end{center}
\end{figure}
}
%%%%%%%%%%%%%%%%%%%%%%%%%%%%%%%%%%%%%%%%%%%%%%%%%%%%%%%%%%%%%%%%%%%%%%%%%%%

\section{Conclusion}
\label{concl}
{We have demonstrated that our models of roAp stars reproduce the observations in great detail. The temperature and luminosity parameters of our best fit are in good agreement with those expected from previous observations \citep{olegbagnulo}, although we find a slightly higher mass. According to \citet{tanya97} the mean magnetic field modulus of $\gamma$ Equ is $\rm \simeq 4$\,kG, which is half of the polar magnetic field strength of 8\,kG as suggested by our best model. At present we can only speculate about a difference of internal magnetic field strengths, to which pulsation is sensitive, relative to surface magnetic fields, which are accessible to  spectroscopy. Furthermore, our models assume a simple, axisymmetric magnetic dipole, which may not sufficiently reflect a more complicated reality. The factor of 2 between (surface) magnetic field modulus and best fitting magnetic pulsation model appears again for the roAp star \object{10\,Aql} \citep{huber}, another {\sc Most} primary target. Certainly, more investigations of this sort have to be conducted in order to solve this problem.

Pulsation amplitude changes for roAp stars were discussed in the literature as a consequence of limited mode life time or beating frequencies. For $\gamma$ Equ we can clearly identify the beating frequencies, which seriously  questions excitation mechanisms for roAp stars based on stochastic processes.

Finally, we could identify the modes of all 7 frequencies detected so far in $\gamma$ Equ  as $\ell = 0$, $1$, $2$ and $4$, with $f_{\rm{6}}$ and $f_{\rm{7}}$ being the harmonics of $f_{\rm{1}}$ produced by non-linear pulsation. 
Since we are not able to calculate non-axisymmetric modes, all of these frequencies are assumed to be ($m=0$) p-modes. While we cannot rule out the possibility of excitation of frequencies with ($m\neq0$), axisymmetric modes are most probable, because of the extremely long rotation period of $\gamma$ Equ and the assumption of a magnetic dipole.
The assignment of any $\ell$-value to these modes, however, has to be understood as a convenient simplification. In the presence of strong magnetic fields a mode's oscillation behavior cannot be described by a single spherical harmonic. Consequently, we alert the reader that analysis techniques using this assumption, e.g. mode identification based on LPVs, should be adjusted for magnetic effects on the pulsation geometry.}

\begin{acknowledgements}
We would like to thank the referee for all the useful comments and suggestions, which greatly helped to improve the quality of this paper. It is also a pleasure to thank M. Cunha, O. Kochukhov, P. Reegen, T. Ryabchikova and H. Shibahashi for valuable discussions. MG, TK and WW have received financial support by the Austrian Fonds zur F\"orderung der wissenschaftlichen Forschung (P17890-N02) and by the Austrian Research Promotion Agency, FFG-ALR. JMM, DBG, AFJM, SR, DS, and GAHW acknowledge funding from the Natural Sciences \& Engineering Research Council (NSERC) Canada. RK is supported by the Canadian Space Agency (CSA).

\end{acknowledgements}

\bibliographystyle{aa}
\bibliography{8830bib}

\end{document}